\documentclass[aps,prl,showpacs,notitlepage,reprint]{revtex4-1}
\usepackage{graphicx}
\usepackage{dcolumn}
\usepackage{amsmath}
\usepackage{bm}
\usepackage{todonotes}
\usepackage[utf8]{inputenc}
\usepackage{amsmath}
\usepackage{natbib}
\usepackage{graphicx}
\usepackage{caption}
\usepackage{subcaption}
\usepackage{lipsum}
\usepackage{float}

\newcommand{\bn}[1]{\mbox{\boldmath$#1$}}

\newcommand{\beq}{\begin{equation}}
\newcommand{\eeq}{\end {equation}}
\newcommand{\bea}{\begin{eqnarray}}
\newcommand{\eea}{\end{eqnarray}}

\begin{document}
\title{Chirality and helicity of optical vortices of small beam waists}

\author{K. Koksal$^{*+}$, {\rm {M.\; Babiker}}}\email{m.babiker@york.ac.uk} 
\affiliation{*Department of Physics, University of York, YO10 5DD, UK} 
\affiliation{$^+$Physics Department, Bitlis Eren University, Bitlis 13000, Turkey}
 \author{V. E. Lembessis}
 \affiliation{Quantum Technology Group, Department of Physics and Astronomy, College of Science, King Saud University, Riyadh 11451, Saudi Arabia}
 \author{J. Yuan}
 \affiliation{Department of Physics, University of York, YO10 5DD, UK} 
\date{\today}

\begin{abstract}
The chirality and helicity of linearly polarised Laguerre-Gaussian (LG) beams are examined.  Such a type of light possesses a substantial longitudinal field when its beam waist is sufficiently small and so gives rise to non-zero chirality and helicity.  In the simplest case of a doughnut beam of winding number $\ell=1$ and another identical to it but has $\ell=-1$, we obtain different chirality and helicity distributions at the focal plane $z=0$. We also show that this chiral behaviour persists and the patterns evolve so that on planes at $z<0$ and $z>0$ the beam convergence contributes differently to the changes in the chirality and helicity distributions. 
\end{abstract}


\maketitle

Recent developments in the generation of laser light beams have highlighted the advances in the ability to focus light to  sub-wavelength dimensions \cite{Dorn2003,Dorn2003a,Bauer2015,Kotlyar2019}.  Optical vortices such as Laguerre-Gaussian (LG) beams of sufficiently small beam waists $w_0$  possess a new feature, namely that the longitudinal (axial) component of the vortex electric field, which is normally regarded as insignificant, now acquires a magnitude comparable to the transverse components.  Furthermore, in such beams the gradients of the  curvature phase and that of the z-dependence of the beam waist $w(z)$ play significant roles in the properties of the light in the vicinity and on either sides of the focal plane.

Some phenomena have been identified as consequences of vortex light of small beam waist, including novel field distributions \cite{Quabis2000,Youngworth2000}, modifications of the spin-orbit interaction \cite{Zhao2007,Tang2010},
 the creation of transverse orbital angular momentum components \cite{Aiello2009,Banzer2013,Aiello2015,Bliokh2015transverse} and changes in the interaction of light with matter \cite{Quinteiro2017}.  The recent experimental report by Wozniak et al \cite{wozniak2018} demonstrated that optical vortex light exhibits a chiral behaviour \cite{Tang2010} in the sense that the vortex nature of the light involves distinguishing between beams with equal but different signs of the winding number $\ell$. In other words it is not possible by rotation alone to superimpose the chirality distribution for $\ell$ on that for  $-\ell$. 
 
This paper is concerned with  the electromagnetic fields of linearly polarised doughnut beams of sufficiently small beam waists.
Our aim is to evaluate the chirality and the helicity distributions, both on and in the vicinity of the focal plane. The evaluations depend on incorporating two ingredients.  The first ingredient involves the inclusion of the longitudinal field components and the second involves taking full account of the convergence phase embodied in the Gouy and curvature  phases and the variations of the beam waist with the axial coordinate.  As will be revealed below, we find that these contribute significantly to the chirality and the helicity  of the beams.

For simplicity we consider the first doughnut modes for which $p=0$ with the same magnitudes but different signs of the winding number $\ell=\pm 1$ and we assume that both doughnut modes are linearly polarised, so neither has optical spin.  For $\ell=\pm 1$ the electric field can be written in cylindrical coordinates ${\bf r}=(\rho,\phi,z)$ in terms of the amplitude function $U_{k10}(\rho,z)$  for $\ell=1$. This amplitude function is exactly the same as $U_{k(-1)0}(\rho,z)$ for $\ell=-1$.  The two phase functions differ, however.  We have for the amplitude functions \cite{babiker2018atoms,koksal2019}
\beq
E^x_{k\ell 0}({\bf r})=U_{k\ell 0}({\rho,z})e^{i\Theta_{k\ell0}({\bf r})}\label{Eee}
\eeq
where $\ell=\pm 1$ and the superscript $x$ in $E^x$ indicates wave polarisation along the x-axis. The amplitude functions for $\ell=\pm 1$ are identical 
\begin{equation}\label{14}
    U_{k\pm 10}(\rho,z)={\cal U}_{k00}\frac{1}{(1+z^2/z_R^2)^{1/2}}\Big( \frac{\rho\sqrt{2}}{w(z)}\Big)e^{-\frac{\rho^2}{w(z)^2}},
\end{equation}
In the above ${\cal U}_{k00}$ is the amplitude for a corresponding plane wave of 
wavevector $k$; $w(z)$ is the beam waist at axial  coordinate $z$ such that 
$w^2(z) = 2(z^2+z_R^2)/kz_R$, where $z_R$ is the Rayleigh range, $w_0=w(0)$ is the beam waist  at the focal plane $z=0$.  The phase functions of the doughnut modes including the convergence phases are as follows
\beq
    \Theta_{k{\pm 1}0}({\bf r})=kz\pm\phi+\theta_{Gouy}+\theta_{curve}
\eeq
where $\theta_{Gouy}$ and $\theta_{curve}$ are given by
\beq
\theta_{Gouy}=-2\tan^{-1}(z/z_{R});\;\;\;\;\theta_{curve}=\frac{k\rho^2z}{2(z^2+z_{R}^2)}\label{converge},
\eeq
 Note that these convergence phases vanish at the focal plane, but have different variations in the planes $z>0$ and $z<0$ on either side of the focal plane.
 
 In addition to the transverse component of the electric field given in Eq.(\ref{Eee}) there must also exist a longitudinal (or axial) component $E^z$.  This is because the total electric field vector ${\bf E}$ must satisfy the transversality condition${\bn {\nabla}}\cdot{\bf E}=0$ where
\beq
{\bf E}=E^x{\bn {\hat x}}+E^z{\bn {\hat z}}
\eeq 
so that the axial component $E^z$ follows formally from the transversality condition and is in a closed analytical form as follows
\begin{equation}
\begin{split}
E^z_{k10}({\bf r})=&-i{\cal U}_{k00}\left(\frac{4 \rho ^2\cos\phi -w(z)^2 e^{-i\phi}}{{\epsilon} k w(z)^4}\right)\\
&\times e^{-\frac{\rho ^2}{w(z)^2}+ i \Theta_{k10}}
\label{vecE2}
\end{split}
\end{equation}
where $\epsilon=\sqrt{k/4 z_R}$. Note that in the derivation leading to the closed analytical expression Eq.(\ref{vecE2}) for the longitudinal electric field we have kept the full z-dependence residing in $w(z)$ using the substitution $w(z) = \sqrt{2(z^2+z_R^2)/kz_R}$ where $z_R=\pi w_0^2/\lambda$ with $w_0$ the beam waist at the focal plane and $\lambda$ the wavelength of the light. We have also taken full account of  the convergence phases $\theta_{Gouy}$ and $\theta_{curve}$.  Similar evaluations for the case $\ell=-1$ have then been carried out straightforwardly leading to the corresponding formalism for the doughnut beam for which the longitudinal electric field is $E^z_{k(-1)0}$.

It is interesting to explore the variations of the magnitude of the longitudinal field $E^z$ for the case $\ell=1$ doughnut mode with different values of $w_0$.   Figure 1 compares the variations of the modulus squared of the longitudinal component $E^z$ with that of the transverse component $E^x$. The variations are considered in the focal plane for two sets of the beam waists.  It is clear from the plots that the longitudinal field is very small for larger $w_0$ and becomes comparable to the transverse field for smaller $w_0$.  It is also easy to deduce from the analytical expression for $E^z$ in Eq.(\ref{vecE2}) on expanding the $\cos\phi$ we immediately show that this longitudinal field component is a superposition of the $\ell=0$ mode and the $\ell=2$ mode. It is clear then that it is the $\ell=0$ mode that accounts for the central peak in Fig.1 (a) and (b).   

\begin{figure}
\begin{subfigure}{.45\textwidth}
  \includegraphics[width=1\linewidth]{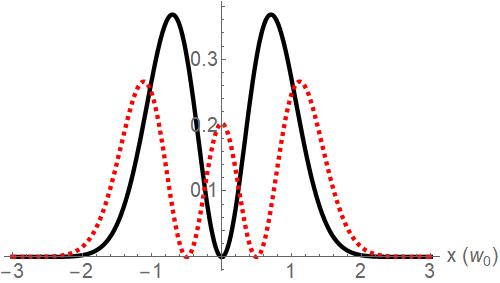}  
  \caption{$w_0=0.5$ $\lambda$}
  \label{fig:sub-first}
\end{subfigure}
\begin{subfigure}{.45\textwidth}
  \includegraphics[width=1\linewidth]{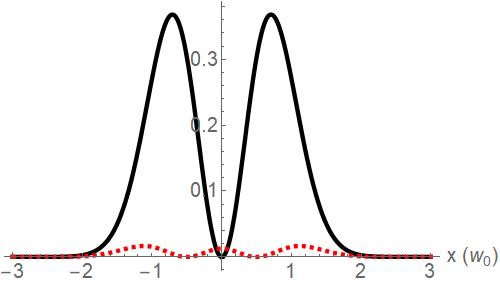}  
  \caption{$w_0=2$ $\lambda$}
  \label{fig:sub-second}
\end{subfigure}
\caption{The in-plane variations of the modulus squared of the electric field components for the $\ell=1, p=0$ Laguerre-Gaussian (doughnut) mode in the focal plane $z=0$. The sub-figures show the corresponding  variations of the longitudinal field $E^z$ (dotted red curve) along with the transverse component $E^x$ (solid black curve). (a) the case of a beam with a small winding number $w_0=0.5\lambda$ where $E^z$ is comparable with $E^x$ and (b) the case of a relatively large beam waist $w_0=2\lambda$ where $E^z$ appears negligibly small in comparison with $E^x$. These results also apply for the case $\ell=-1, p=0$}
\end{figure}

The general definitions of the chirality and the helicity of a light field require, in addition to the usual electromagnetic fields,  the introduction of dual fields.  Accounts of optical chirality and helicity can be found in  \cite{inproceedings,inproceedings2} along with the recent literature on this subject. In the Coulomb gauge we have for the chirality
\beq
\chi=\frac{1}{2}\left(-\epsilon_0{\bf E}\cdot{\dot{\bf B}}+{\bf B}\cdot{\dot {\bf D}}\right)
\eeq
and for the helicity
\beq
\eta=\frac{1}{2}\left(\sqrt{\frac{\epsilon_0}{\mu_0}}{\bf A}\cdot{\bf B}-\sqrt{\frac{\mu_0}{\epsilon_0}}{\bf C}\cdot{\bf D}\right)
\eeq
where ${\bf D}=\epsilon_0{\bf E}$ is the displacement field while  ${\bf C}$ is a second vector potential dual to the usual vector potential ${\bf A}$ such that ${\bn {\nabla}}\times{\bf C}=-{\bf D}$ and ${\bn {\dot C}}=-{\bf B}/\mu_0$.
However, we will assume that we are dealing with cycle-averaged monochromatic fields in which case the the cycle-averaged chirality ${\bar {\chi}}$ and the helicity   ${\bar {\eta}}$ are given by
\bea
{\bar {\chi}}&=&-\frac{\omega\epsilon_0}{2}\Im[{{\bf E}^*\cdot{\bf B}}]\nonumber\\
&=&\frac{\omega^2}{c}{\bar {\eta}}
\label{chieta}
\eea
Thus in order to evaluate the helicity and the chirality we need to determine the magnetic field components using Maxwell's equation ${\bn {\nabla}}\times {\bf E}=i\omega{\bf B}$. There are three components of the magnetic field, given by
\beq
{\bf B}=\frac{1}{i\omega}\left\{\partial_yE^z{\bn {\hat x}}+(\partial_zE^x-\partial_xE^z){\bn {\hat y}}-\partial_yE^x{\bn {\hat z}}\right\}
\eeq
However, since $E^y=0$, then as far as the evaluation the dot product in Eq.(\ref{chieta}) is concerned only the x and z components of the magnetic field are relevant and these are given by 
\begin{widetext}
\begin{equation}
\begin{split}
B^x_{k10}=&{\cal U}_{k00}\frac{2 \rho  z_R \left(\sin 2 \phi  \left(\left(3 z_R-i z\right) w(z)^2+4 i \rho ^2 \left(z+i z_R\right)\right)+\left(z+3 i z_R\right) w(z)^2 \cos 2 \phi -\left(z-i z_R\right) w(z)^2\right)}{\omega  w(z)^6 \left(k z_R\right){}^{3/2}}\\
&\times e^{-\frac{\rho^2}{w(z)^2}+i\Theta_{k10}}
\end{split}
\end{equation}
\beq
B^z_{k10}={\cal U}_{k00}e^{-\frac{\rho^2}{w(z)^2}}e^{i\Theta_{k10}}\frac{-2 i z_R w(z)^2 \sin (\phi )+2 z_R w(z)^2 \cos (\phi )+4 \rho ^2 \left(z+i z_R\right) \sin (\phi )}{\omega  w(z)^4 \sqrt{k z_R}}
\eeq
\end{widetext}
Similar evaluations have been carried out straightforwardly for the magnetic fields in the case of $\ell=-1$. The results for the chirality distribution in the focal plane are shown in Fig.2 in which the two doughnut beams have the same waist $w_0=0.5$ $\lambda$ and differ only in the sign of the winding number. Figure 2(a) concerns the doughnut beam with winding number $\ell=1$ and 2(b) with negative winding number $\ell=-1$. The differences are clear in that regions of the focal plane where the chirality is high (bright red) in 2(a) are replaced by regions of low chirality (in blue) in 2(b) and vise versa.  The distributions cannot be described as mirror image of each other.  These  results confirm the chiral character of the vortex light that is linearly polarised and so has no spin. 

Note that the assumption that we are dealing with cycle averaged fields meant that the helicity is directly proportional to the chirality, as described by Eq. (\ref{chieta}).

\begin{figure}
\begin{subfigure}{.45\textwidth}
  \centering
  \includegraphics[width=.7\linewidth]{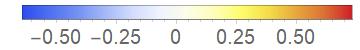}  
  \label{fig:legend}
\end{subfigure}
\begin{subfigure}{.45\textwidth}
  \centering
  \includegraphics[width=.7\linewidth]{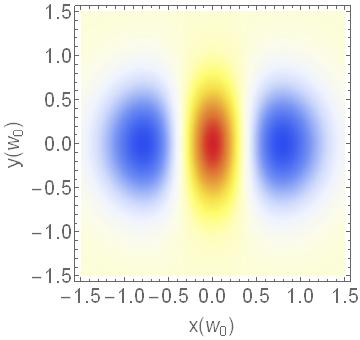}  
  \caption{$\ell=1$ }
  \label{fig:sub-firsta}
\end{subfigure}

\begin{subfigure}{.45\textwidth}
  \centering
  \includegraphics[width=.7\linewidth]{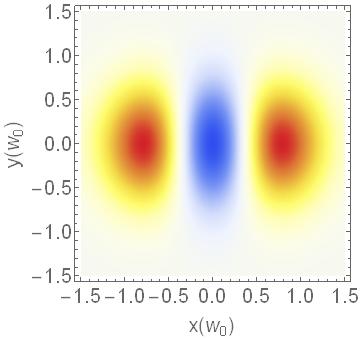}  
  \caption{$\ell=-1$}
  \label{fig:sub-seconda}
\end{subfigure}
\caption{The chirality distribution in the focal plane $z=0$ for two doughnut beams for which the waist is $w_0=0.5 \lambda$ (a) the case of a doughnut beam with winding number $\ell=1$ and (b) the case with negative winding number $\ell=-1$.}
\label{fig:helz0}
\end{figure}
The consideration so far have been concerned with the properties of the vortex beam in the focal plane $z=0$.  It is, however, of interest to explore what changes occur in planes to the left $z<0$ and to the right $z>0$ of the focal plane to ascertain that the chirality features continue beyond the focal plane.  The general expressions displayed above for the electric and magnetic fields show explicit spatial dependence on the axial and radial coordinates in both their amplitudes and phases, including those arising from the presence of the longitudinal field component.  The convergence phase comprising the Gouy phase $\theta_{Gouy}$ and curvature phase $\theta_{curv}$ are both operative as displayed in Fig. 3 as well as the variations of $w(z)$.  These phase functions can have relatively large gradients in the case of small beam waists in all planes in the vicinity of the focal plane. 

 \begin{figure}
  \includegraphics[width=.8\linewidth]{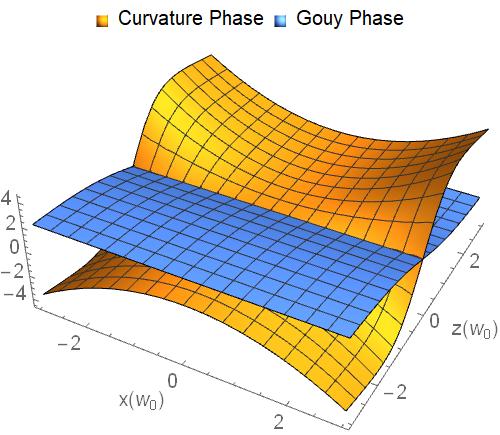}  
  \caption{Variations in the $xz$-plane of the Gouy phase $\theta_{Gouy}$ and the curvature phase $\theta_{curve}$ of the doughnut beam for which $w_0=0.5\lambda$.  Note the sign change across the plane $z=0$.}
  \label{fig:gouy}
\end{figure}
 
Direct evaluations of the chirality distributions on the planes $z>0=+2w_0$ and $z<0=-2w_0$ (as done for the focal plane in Fig.2) leads to the rather different distributions shown in Fig.4. These results demonstrate that the chirality features exhibited in the focal plane still exist on both sides of the focal plane, but the distributions have evolved from their forms in the focal plane.  In Figs.4 (a and b) which concerns the plane $z=+2w_0$ the same chirality feature persists between the case $\ell=1$ and the case $\ell=-1$  but both distributions are twisted by equal and opposite angles in comparison with the case in the focal plane in Fig.2. Figures 4( c and d) display the chirality distributions in the plane $z=-2w_0$ to the left of the focal plane and, once again, for $\ell=1$ and $\ell=-1$.  As before, there is a twist of the patterns relative to those in the focal plane, but the twisting is opposite to those for the same $\ell$ but on equidistant planes on different sides of the focal plane. In fact Fig. 4(a) and 4(d) show the same twist but reciprocal intensity distributions and the same features are displayed between Fig.4(b) and 4(c). 
\begin{figure*}[ht]
\begin{subfigure}{.8\columnwidth}
  \centering
  \includegraphics[width=1\columnwidth]{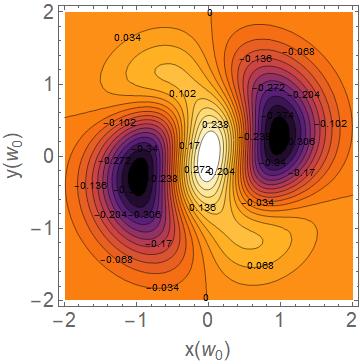}  
  \caption{$\ell=1$ and $z=2w_0$}
  \label{fig:sub-firstb}
\end{subfigure}
\begin{subfigure}{.8\columnwidth}
  \centering
  \includegraphics[width=1\columnwidth]{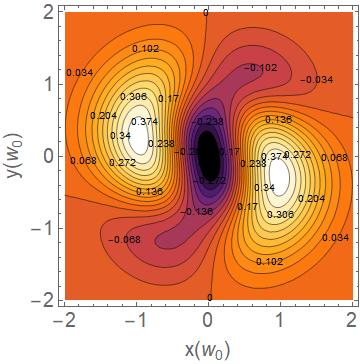}  
  \caption{$\ell=-1$ and $z=2w_0$}
  \label{fig:sub-secondb}
\end{subfigure}


\begin{subfigure}{.8\columnwidth}
  \centering
  \includegraphics[width=1\columnwidth]{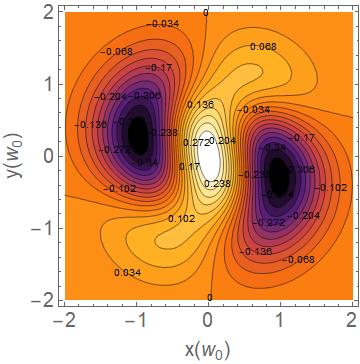}  
  \caption{$\ell=1$ and $z=-2w_0$}
  \label{fig:sub-firstc}
\end{subfigure}
\begin{subfigure}{.8\columnwidth}
  \centering
  \includegraphics[width=1\columnwidth]{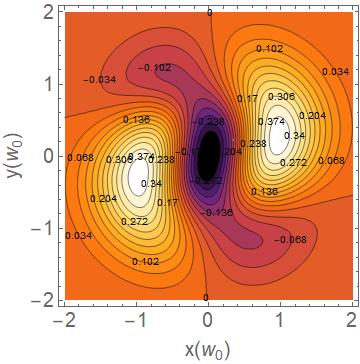}  
  \caption{$\ell=-1$ and $z=-2w_0$}
  \label{fig:sub-secondc}
\end{subfigure}
\caption{The chirality distribution in the plane $z=+2w_0$ and $z=-2w_0$ in the vicinity of the focal plane for two  doughnut beams for which the waist is $w_0=0.5\lambda$: Left panel (a and c): the case of a doughnut beam with winding number $\ell=1$ (a) plane $z==+2w_0$ and (b) plane $z=-2w_0$. The right panel (b and d) is the same as the left panel, but for negative winding number $\ell=-1$.}
\label{fig:fi43}
\end{figure*}

In conclusion, we have examined two of the main properties of optical vortices, namely chirality and helicity and have shown that in the regime where the beams have small beam waist $w_0$, the longitudinal component of the electric field and its associated magnetic field become of the same order of magnitude as those of the corresponding transverse components.  Such longitudinal components  are normally negligible for beams with relatively large beam waist $w_0$.  Our work has confirmed that vortex light in the form of doughnut beams exhibit chirality in the sense that a change  of the sign of the winding number $\ell$ produces a different chirality distribution.  We have also shown that the convergence phases give rise to features additional to chirality in the vicinity of the focal plane in the form of a rotation of patterns and changes in equidistant planes on either sides of the focal plane.

The chiral nature of the vortex light has also been investigated experimentally  by Wozniak et al \cite{wozniak2018} and their experimental findings were shown to conform with the results arising from the vectorial diffraction theory (see also \cite{Zhao2007} and references therein).  The Wozniak et al's work had involved passing each doughnut beam through a lens and the chirality they measured was due to the fields observed in the back focal plane of this lens. With this arrangement they, too, confirmed the chiral character in the case of the two linearly polarized LG beams subject to focusing by the lens, one with winding number $\ell=1$ and another with winding number $\ell=-1$ with neither beam possessing spin angular momentum. However, their chirality distribution in the back focal plane of the lens differs from our results which have been produced in the absence of a lens as shown in Figs. 2 and 4.  We suggest that the differences lie in the manner in which the doughnut beams responded to the focusing lens in \cite{wozniak2018}.  The changes could be related to the effect of the focusing of the linearly polarised vortex light, as pointed out in \cite{Dorn2003a} and it has also been suggested that linearly polarised light falling onto the back focal plane of the lens becomes uniformly polarized \cite{levy2019}. These scenarios are not applicable to what we have been concerned with in this Letter in which we have dealt with linearly-polarised doughnut beams with small $w_0$ and the chirality and helicity features we have confirmed are applicable to such beams subject to no further focusing by a lens as was the case investigated in \cite{wozniak2018}.  

To summarise, in this paper we have  derived analytical expressions for  the longitudinal electric field component  and the corresponding magnetic field components  for doughnut modes and shown that these are comparable in magnitude to the transverse components, but  only for small $w_0$. We have also shown that no meaningful chirality and helicity properties exist without the substantial longitudinal components and demonstrated explicitly the chirality distributions in the focal plane. We have also explored the role of the Gouy and curvature phase gradients in the shape of the chirality density distributions on different sides of the focal plane. We have concentrated on generating figures mainly on cycle-averaged chirality density.  This is because under such conditions,  apart from a constant factor, the helicity distribution plots  would look exactly the same as the chirality distributions.  We have highlighted the interesting dual symmetry roles of the sign of the winding number and the sign of  the plane side (to the left and right of the focal plane). We have explained the meaning of chirality in relation to Fig. 2 evaluated in the focal plane $z=0$ where the Gouy and curvature phases vanish and so have no role to play as regards the chirality on the focal plane. We interpreted the interesting chirality shapes in Fig. 4 where $z\neq 0$ and emphasised the role played by the curvature and Gouy phases in the chirality distributions on planes to the left and to the right of the focal plane. We explained how changing the sign of the winding number and the signs of the left and right planes relative to the focal plane give rise to similarities and differences in the chirality distributions

\section*{Acknowledgements}
We are grateful to Professor S. M.  Barnett for helpful discussions. KK wishes to thank Tubitak and Bitlis Eren University for financial support (under the project:BEBAP 2017.19) during his sabbatical year at the University of York where this work was initiated.
\bibliography{references}
\end{document}